\documentclass[journal=jacsat,manuscript=article]{achemso}

\usepackage[version=3]{mhchem} 
\usepackage{xcolor}
\usepackage{siunitx}
\usepackage{multirow}
\usepackage{amssymb}

\author{John J. Karnes}
\email{karnes@llnl.gov}
 \affiliation{%
 {Lawrence Livermore National Laboratory
       Livermore, California 94550, United States}
}
\author{Andrew J. Pascall}
 \affiliation{%
 {Lawrence Livermore National Laboratory
       Livermore, California 94550, United States}
}
\author{Christoph Rehbock}
 \affiliation{%
 {Institute of Technical Chemistry I, University of Duisburg-Essen and Center for Nanointegration Duisburg-Essen (CENIDE), 45141 Essen, Germany}
}
\author{Vaijayanthi Ramesh}
 \affiliation{%
 {Institute of Technical Chemistry I, University of Duisburg-Essen and Center for Nanointegration Duisburg-Essen (CENIDE), 45141 Essen, Germany}
}
\author{Marcus A. Worsley}
 \affiliation{%
 {Lawrence Livermore National Laboratory
       Livermore, California 94550, United States}
}
\author{Stephan Barcikowski}
 \affiliation{%
 {Institute of Technical Chemistry I, University of Duisburg-Essen and Center for Nanointegration Duisburg-Essen (CENIDE), 45141 Essen, Germany}
}
\author{Elaine Lee}
 \affiliation{%
 {Lawrence Livermore National Laboratory
       Livermore, California 94550, United States}
}
\author{Brian Giera}
\email{giera1@llnl.gov}
 \affiliation{%
 {Lawrence Livermore National Laboratory
       Livermore, California 94550, United States}
}

\title{Particle-Based Simulations of Electrophoretic Deposition with Adaptive Physics Models} 

\abbreviations{IR,NMR,UV}

\begin{document}

\begin{abstract}
  This work represents an extension of mesoscale particle-based modeling of electrophoretic deposition (EPD), which has relied exclusively on pairwise interparticle interactions described by Derjaguin-Landau-Verwey-Overbeek (DLVO) theory. With this standard treatment, particles continuously move and interact via excluded volume and electrostatic pair potentials under the influence of external fields throughout the EPD process. The physics imposed by DLVO theory may not be appropriate to describe all systems, considering the vast material, operational, and application space available to EPD. As such, we present three modifications to standard particle-based models, each rooted in the ability to dynamically change interparticle interactions as simulated deposition progresses. This approach allows simulations to capture charge transfer and/or irreversible adsorption based on tunable parameters. We evaluate and compare simulated deposits formed under new physical assumptions, demonstrating the range of systems that these adaptive physics models may capture.  
\end{abstract}


\section{Introduction}

Electrophoretic deposition (EPD) is a well-studied process where colloidal particles, driven by an electric field, accumulate at an electrode to form a deposit. The accessible property space of these resulting deposits is enormous due to the large number of independently tunable parameters governing EPD, such as the electrode substrate~\cite{}, electric field strength~\cite{} and waveform~\cite{}, particle morphology~\cite{} and loading~\cite{}, supernatant characteristics~\cite{}, and so on. As such, EPD is found in a wide variety of applications, ranging from neural electrodes~\cite{ramesh2021comparing,osti_1860812}, biomaterials processing~\cite{Boccaccini2010,AVCU201969,SIKKEMA2020102272}, high-contrast low-voltage~\cite{PhysRevMaterials.4.075802,Giera_2018} and quantum dot-based displays~\cite{Zhao2021}, battery~\cite{https://doi.org/10.1002/batt.201900017} and supercapacitor~\cite{NGUYEN2019122150} manufacturing, fog harvesting~\cite{GHOSH20203777}, superhydrophobic~\cite{Saji2021} and energetic~\cite{Sullivan2012} coatings, etc.  Though efforts to optimize EPD deposits must be application-specific, the community typically relies on a general set of long-standing models for the stability of colloids and their deposition. For instance, particles are assumed to interact strictly according to Derjaguin-Landau-Verwey-Overbeek (DLVO) theory and adhere to a deposition rate that scales linearly with the strength of a uniformly applied field~\cite{hamaker1940formation}. 

Fortunately, numerous models exist to accommodate the nuances of actual EPD systems. Ferrarri, et. al.'s extensive review~\cite{FERRARI20101069} lays out several analytical expressions that account for commonly encountered EPD scenarios, including flat~\cite{hamaker1940formation} and cylindrical electrodes~\cite{biesheuvel1999theory} with variable~\cite{sarkar1996electrophoretic} and/or highly concentrated~\cite{https://doi.org/10.1111/j.1151-2916.1999.tb01939.x} particle loading. Empirically validated models exist that account for suspension resistivity variations alone~\cite{anne2005mathematical} and in combination with solid loading~\cite{ferrari2006resistivity}. Several models rely on finite element analysis (FEA)  to simulate  coating at the industrial-scale~\cite{VERMA2020101075} and deposit thickness variations that result from electrode edge effects~\cite{Pascall2012}. A recent multi-physics FEA model accounts for transient phenomena arising in a specialized EPD cell with cross flow across empirical time and length scales. It reproduces actual deposit morphologies over a wide range of inlet flow rates and applied voltages~\cite{SALAZARDETROYA2021110000}. Finite element based approaches offer unique insights, specifically into how the deposition thickness evolves under (possibly) complicated electrode geometries and electric, flow, and particle concentration fields.

Particle-based models (PBMs) elucidate how deposit \emph{microstructure} depends on the interplay between inter-particle and electric field forces during deposition. Specifically, PBMs compute the trajectory of all colloidal particles in a simulated control volume using Newton's equations of motion and assumed interaction potentials that model particle-particle interactions in a suspension bounded by a charged electrode. As such, PBMs reveal how particles accumulate at the electrode, though typically over shorter time scales than FEA-based simulations due to greater computational expense. PBM simulations accurately reproduce analytical models of bulk electrophoretic motion across an expansive experimentally relevant regime~\cite{Giera2015} and have elucidated how the degree of ordering within multi-layed deposits depends on \emph{both} ionic concentration and field strength~\cite{gieraMesoscaleParticleBasedModel2017}. The same PBM applied to sub-monolayer deposits revealed correlations between  electrochemical properties and composition of the supernatant with coating homogeneity~\cite{osti_1860812}. PBM particle-spacing predictions were experimentally validated using a novel EPD cell that enables \emph{in situ} characterization. Considering long-standing critiques of DLVO~\cite{mcbride1997critique} applied to the wide range of available colloidal suspensions, it is fortunate that the underlying inter-particle potential computed by PBMs is fully customizable. 

\begin{figure}[h]
\includegraphics[width=4.5in]{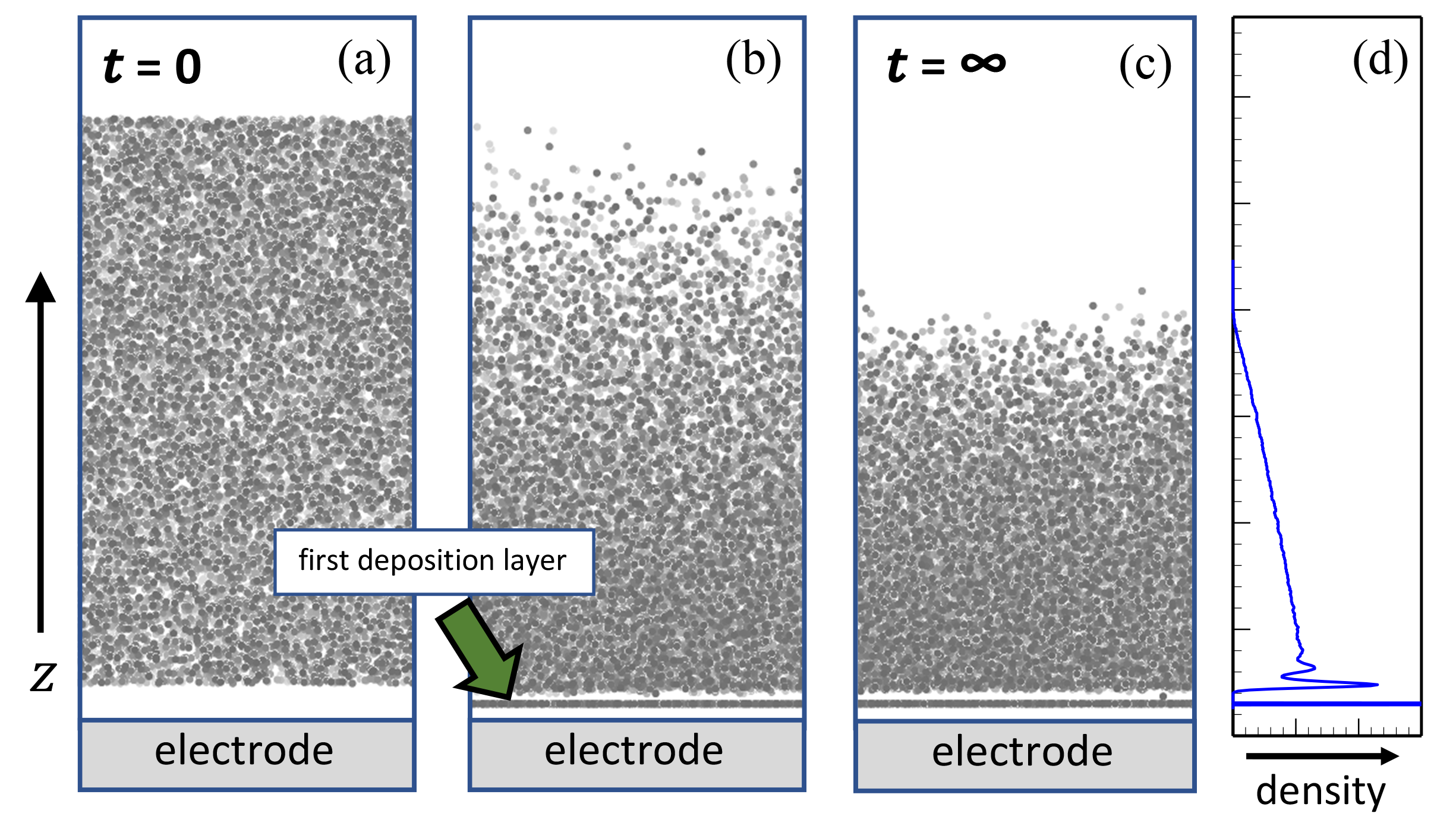}
\caption{\label{fig:epd-toon} Simulation snapshots show typical particle configurations for a conventional DLVO particle-based EPD simulation at (a) the start of the simulation, (b) as deposition begins, and (c) at equilibrium. (d) Equilibrium density profiles (with vertical axis aligned with the simulation $z$-coordinate) reveal a maximum density at the electrode followed by a few layers of decaying order and a region of deposited material whose density decreases linearly toward the supernatant region devoid of particles.}
\end{figure}

To date, PBMs exclusively assumed that interparticle interactions are of a fixed functional form that follows DLVO theory. Figure \ref{fig:epd-toon} shows representative simulation snapshots from a particle-based EPD simulation. Panel (a) shows the starting configuration, which consists of randomly distributed colloid particles. When the external electric field is applied (panel (b)), the particles begin to migrate in the $-z$ direction, toward the electrode, assembling into a deposit. Panel (c) shows the representative simulation at equilibrium. We note that the position of each particle is stored and represented as a point, not a space-filling sphere, in these simulations and the data shown in this work. This representation allows us to clearly observe surface-induced ordering of the deposits onto the geometrically flat electrode in the snapshots by illustrating each particle as a gray point whose radius is smaller than the particle's physical radius. The apparent `space' between the first deposition layer and the electrode (and the first layer and the bulk deposit) emerges from and is enhanced by this representation. Gray rectangles are included in this overview figure to indicate the position of the electrode. Subsequent simulation snapshots show only the particles. The density profile in panel (d) corresponds to the equilibrium configuration in (c) and quantifies the surface-induced ordering. We again note that this data captures the center points of the particles. Density profiles specifically report the density of the particles' centers, not a bulk density, which enhances our ability to observe ordering of particles in the deposit. 

In this conventional PBM approach, interparticle interactions and colloidal motions described by an unchanging physics model persist at all times during deposition. Given the diversity of colloidal suspension and deposition conditions, we argue that many EPD systems of interest exhibit behavior not described by a static set of pairwise interactions defined a priori. Thus, we pose a PBM  framework that allows for exploration of particles with physics models that may change on-the-fly according to a set of pre-defined heuristics. With these new models, we simulated parameter sweeps that vary the applied electric field and characteristic interparticle electrostatic repulsion. Specifically we simulate conditions that can occur in real systems in which particles can discharge and/or become immobile upon deposition e.g. when a metal nanoparticle deposits on a metal surface and forms a covalent or metallic bond at the interface. It is beyond the scope of this work to explore all parameter space made accessible by these new simulation protocols. Nevertheless, we simulate experimentally-accessible EPD conditions and compare particle packing density and degree of ordering resulting from the four model types. Further, the physical parameters used in these simulations may be adjusted to explore any EPD system of interest.



\section{Model and simulation details}

\subsection{Interaction potentials}

Mesoscale particle dynamics simulations are based on the work of Giera and collaborators.~\cite{gieraMesoscaleParticleBasedModel2017} We refer the reader to this work for a full description of interaction potentials and model parameterization, but summarize the components most salient to the present work here.

The total energy acting on a colloid particle is a combination of the pairwise sum of solvent-colloid, colloid-colloid, and system-colloid interactions,
\begin{equation}
  U(\textbf{r}^{N_{\text{col}}})=\sum_{i<j}^{N_{\text{col}}} \bigl[U_{\text{DLVO}}(r)+ U_{\text{solvent}}(r) \bigr] +  \sum_{i=1}^{N_{\text{col}}} \bigl[ U_{\text{wall}}(z)+U_{\text{e-field}}(z) \bigr] 
\end{equation}\label{totalE}

\noindent where $r$ is the distance between an $i$-$j$ colloid pair and $z$ is a colloid's distance from the electrode, here represented by a flat, uncharged plane in the $x$-$y$ plane. The solvent is treated implicitly by adding pairwise dissipative lubrication and Brownian forces.~\cite{ballSimulationTechniqueMany1997,kumarOriginsAnomalousStress2010,ermakBrownianDynamicsHydrodynamic1978,kumarMicroscaleDynamicsSuspensions2010,bybeeHydrodynamicSimulationsColloidal2009}

Pairwise colloidal interactions are treated with the DLVO potential, 
\begin{equation}
  U_{\text{DLVO}}(r)=U_{\text{Yuk}}(r)+U_{\text{steric}}(r)+U_{\text{vdW-a}}(r)
  \label{eqn:dlvo}
\end{equation}

\noindent which represents the sum of screened electrostatic repulsions by a variant of the Yukawa potential,~\cite{safranStatisticalThermodynamicsSurfaces2019} $U_{\text{Yuk}}$, and captures steric repulsion and van der Waals attractions separately~\cite{everaersInteractionPotentialsSoft2003} as detailed in Reference \citenum{gieraMesoscaleParticleBasedModel2017}. Since this work is inspired by the deposition of platinum nanoparticles we use a different value for the Hamaker constant than Reference \citenum{gieraMesoscaleParticleBasedModel2017}, where the model was based on polystyrene particles: $A_\text{col}=10^{-20}$ J in this work versus $A_\text{col}=10^{-21}$ J in Reference \citenum{gieraMesoscaleParticleBasedModel2017}.\cite{prieve_simplified_1988,schunk_performance_2012} The Yukawa potential has an exponential form, 
	$U_{\rm Yuk}(r) \propto \exp \left ( -r/ \lambda_{\rm D} \right ) $
, in which the Debye length ($\lambda_D$) is the characteristic interparticle distance where electrostatic repulsion between particles is important. In this work we consider several values of $\lambda_D$, defined as 
\begin{equation}
  \lambda_D=\sqrt{\frac{\epsilon \epsilon_0 k_B T}{(qe)^2 \rho_{\pm}}}
  \label{eqn:lamd}
\end{equation}

\noindent where $\epsilon$ is the relative permittivity of the solvent, $\epsilon_0$ is the permittivity of free space, $k_B$ is Boltzmann's constant, and $T$ is temperature. We note that this definition of $\lambda_D$ holds for colloidal suspensions in a symmetrical electrolyte with $\lvert q_{\pm} \rvert=q$, $e$ is the elementary charge, and $\rho_{\pm}$ is the bulk concentration of ions. As Eq.~\ref{eqn:lamd} indicates, the electrostatic repulsion between colloids is easy to tune empirically, especially in aqueous suspensions, e.g.  $\lambda_D$ increases as $\rho_{\pm}$ decreases via dilution, \emph{vice versa}. For our particle-based simulations, the computational expense scales with the number of pairwise energies with $U_{\rm DLVO}(r) \sim k_{B}T$, which increases with $\lambda_D$.

Interactions of the colloid with the electrode, $U_{\text{wall}}$, and the electric field, $U_{\text{e-field}}$ are single body potentials, i.e., a function of the particles' $z$-position. The colloid-wall interaction prevents the particles from moving through the simulation cell wall in the $-z$-direction, resulting in simulated deposition. Electrophoretic motion is simulated by application of a uniform, external electric field
\begin{equation}
  U_{\text{e-field}}(z)=E_{\text{field}}q_{\text{col}}z
  \label{eqn:wall}
\end{equation}

\noindent where $E_{\text{field}}$ is the strength of the electric field and $q_{\text{col}}$ is the effective charge on a colloid particle in the direction perpendicular to the electrode, obtained from a force balance on a colloid particle experiencing steady-state hydrodynamics drag in an electric field.~\cite{probsteinPhysicochemicalHydrodynamicsIntroduction2013}

\subsection{Alternative deposition models}

As described earlier, we introduce two new physical scenarios that may be employed individually or simultaneously to more accurately simulate a wide range of EPD applications. 

\begin{figure}[h!!!]
\includegraphics[width=3.in]{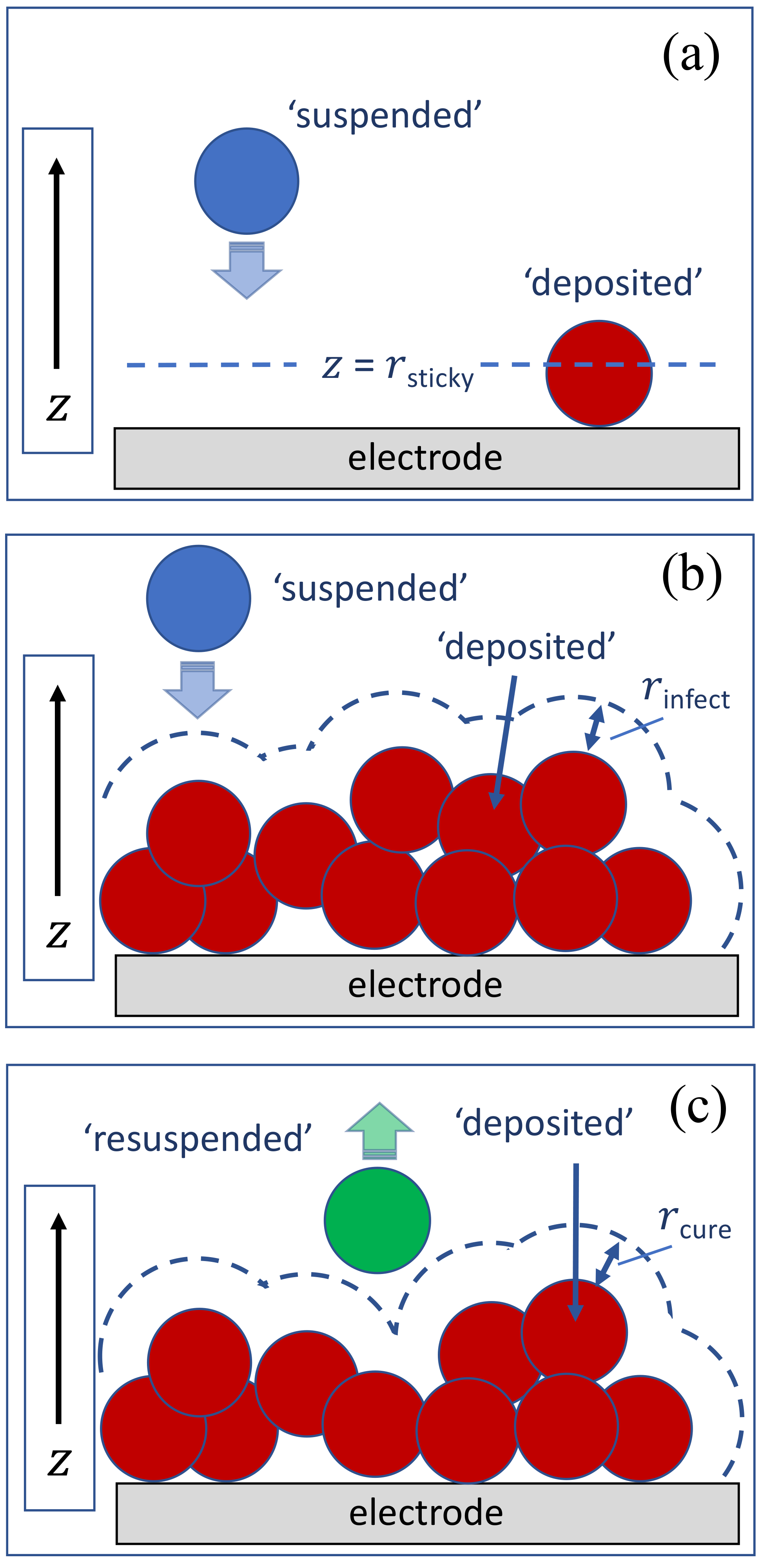}
\caption{\label{fig:fix-toons} Cartoon schematics of the three schemes used to simulate alternative deposition models. (a) In ``sticky'' a suspended particle is flagged as deposited when in close proximity to the electrode. (b) ``Infect'' permits the propagation of the `deposited' state to an arbitrary distance away from the electrode. (c) ``Cure'' flags particles as resuspended when they separate from the deposit }
\end{figure}

In the first scenario, \textit{immobile}, we consider systems where particle mobility decreases after deposition and model the extreme case of completely immobile deposited particles, e.g. when practically irreversible chemical bonds between particle and wall or particle and particle form upon deposition. Our implementation consists of two parts. The first tracks each particle-electrode distance $r_i$ (equivalent to $z_i$ in these simulations). When a given particle is within a cutoff distance ($r_{\text{sticky}}$) of the electrode, the particle is flagged as deposited and prevented from moving by removing it from the molecular dynamics integrator. Though this renders the particle immobile, the particle continues to be accounted for in all interaction potential calculations; the particle is only rendered immobile. The second component of this implementation accounts for similar deposition beyond the first colloid layer, necessary since all deposited particles beyond the first layer would have $z_i>r_{\text{sticky}}$. To handle this deposition, we track the separation distances of all deposited-suspended $i$-$j$ pairs during the simulation. If any mobile colloid particle is within $r_{\text{infect}}$ of a deposited particle, the mobile particle is flagged as deposited and similarly removed from the integrator. Figure \ref{fig:fix-toons} (a) and (b) are schematics of this two part implementation.

Another physically-relevant case not captured by existing PBMs is when a particle's charge changes upon deposition due to being in intimate contact with the electrode, e.g. when a conductive contact between a metal particle and a metal electrode is established. This case includes particles deposited directly onto the electrode and particles that come into contact with the accumulating deposit, which is in contact with the electrode. Our proof-of-concept simulations capture this by making deposited particles assume the electrostatic interaction behavior of the electrode, in practice by zeroing the Yukawa potential between deposited and non-deposited particles. We refer to this scenario as \textit{decharge}. This change to the newly-deposited particle's force field is made when the particle-electrode distance $r_i$ is less than a designated cutoff distance, similar in implementation to the particle immobilization detailed above. Also similar to the immobilization scheme, particles tagged as deposited can `infect' other particles that move within a cutoff distance $r_{\text{infect}}$ of an altered particle. These other particles are then tagged as deposited and their force fields are modified accordingly. These implementations are summarized by the schematics in Figure \ref{fig:fix-toons}(a) and (b). However, because these `infected' particles may be mobile, additional considerations must be made for the case of desorption or resuspension. If no further adjustments were made, a rogue modified particle could become resuspended, diffuse into the bulk, and propagate its electrode-like `decharge' electrostatics into the bulk suspension. To counter this non-physical behavior we add an additional state, `resuspended,' where the `infected' particle retains the altered electrostatics but loses the ability to change the electrostatics of other particles. This third scheme, shown in Figure \ref{fig:fix-toons} (c), flags the deposited particle as resuspended if it is not within a cutoff distance $r_{\text{cure}}$ of either the electrode or other deposited particles. 


The three schemes in Figure \ref{fig:fix-toons} were implemented as subroutines (`fixes') compatible with the 19-March-2020 release of the open source molecular dynamics environment LAMMPS.~\cite{plimptonFastParallelAlgorithms1995,thompsonLAMMPSFlexibleSimulation2021} Source code for these schemes, named \texttt{fix sticky}, \texttt{fix infect}, and \texttt{fix cure}, is included in the Supporting Information.

In practice, combinations of the three implementations shown in Figure \ref{fig:fix-toons}, \textit{sticky}, \textit{infect}, and \textit{cure}, are applied to modify the physics models of colloid particles on-the-fly, enabling the \textit{immobile} and/or \textit{decharge} scenarios. Table \ref{tbl:fixes} summarizes the changes to `deposited' colloids when using any permutation of these alternative deposition models. For the remainder of this work, we use the nomenclature \textit{standard}/\textit{decharge} and \textit{mobile}/\textit{immobile} to indicate the \textit{off}/\textit{on} states of the two scenarios. 

\begin{table}
  \caption{Alternative deposition models } 
  \label{tbl:fixes}
  \begin{center}
  \renewcommand\arraystretch{1.3}
  \begin{tabular}{|c|c|c|}
    \hline
    \multirow{2}{*}{{\bf Model Type}} & \multicolumn{2}{c|}{if $i$ = `deposited'...}\\
    \cline{2-3}
         & ...$U_\text{Yuk}=0 ?$ & ...immobilize?\\
    \hline
    standard/mobile & no & no\\
    decharge/mobile & yes & no\\
    decharge/immobile & yes & yes\\
    standard/immobile & no & yes\\
    \hline
  \end{tabular}
  \end{center}
\end{table}

\subsection{Simulation parameters}

The proof-of-principle simulations presented in this work model a system of 10 nm diameter spherical platinum particles in an aqueous suspension, as in on our earlier studies of neural electrode coatings.~\cite{ramesh2021comparing,ramesh_electrophoretic_2022} Table \ref{tbl:param} summarizes the physical parameters used to specify these simulations. We note that two parameters of particular interest to EPD researchers, Debye length ($\lambda_D$) and electric field strength ($E_z$), were varied in our simulations. Three values of $\lambda_D$, 0.5, 5.0, and 50.0 nm, and three values of $E_z$, 10, 50, and 250 kV/m, were used. The Yukawa potential for each respective $\lambda_D$ was implemented with a smooth cutoff at the distance where $U_\text{Yuk}$ was less than 1\% of $k_BT$. See the Supporting Information for the determination of Yukawa potential cutoff distances. Varying $\lambda_D$ and $E_z$ resulted in 9 parameter-based permutations and 4 deposition model permutations: a total of 36 total EPD simulations. 

Each simulation consisted of 25,920 particles in a $720\times720\times550$ nm box at a bulk volume fraction of $\Phi_{\rm bulk} = $0.05. In this bulk arrangement, the mean pairwise spacing of the particles equals $\bar{r}_{\rm bulk} = r/\sqrt[3]{\Phi_{\rm bulk}}$. Starting configurations were generated by sequentially assigning randomized positions to particles whose pairwise distances are $\ge 0.7 \bar{r}_{\rm bulk}$. Though smaller initial interparticle spacings are physically permissible, this approach led to the most robust performance of the LAMMPS integrator. The simulation cells are periodic in $x$ and $y$ with the short dimension ($z$) normal to the electrode. The periodic dimensions for all simulations were set to be twice the longest force cutoff distance, $\text{Yuk}^\text{cutoff}(\lambda_D=50\text{ nm}) = 360$~nm.

\begin{table}
  \caption{Parameters used in EPD simulations.} 
  \label{tbl:param}
  \begin{tabular}{ccc}
    \hline
    description & symbol & value(s) \\
    \hline

    particle density & $\rho$ & \SI{21}{g/cm^3}\\
    particle radius & $r$ & \SI{5.0}{nm}\\
    number of particles & $N_\text{col}$ & 25920\\
    bulk volume fraction & $\Phi_\text{bulk}$ & 0.05\\
    zeta potential & $\zeta$ & \SI{-0.050}{V}\\
    Hamaker constant\cite{prieve_simplified_1988,schunk_performance_2012} & $A_\text{col}$ & $10^{-20}$ J\\
    relative permittivity & $\epsilon$ & 78.38\\
    dynamic viscosity & $\eta$ & \SI{e-3}{\pascal\second}\\
    temperature & $T$ & \SI{300}{K}\\
    Debye length & $\lambda_D$ & 0.5, 5, 50 nm\\
    electric field strength & $E_z$ & 10, 50, 250 kV/m\\
    electrode deposition cutoff & $r_\text{sticky}$ & 2.0 $r$ \\
    deposition proximity cutoff & $r_\text{infect}$ & 1.2 $r$ \\
    resuspension proximity cutoff & $r_\text{cure}$ & 1.4 $r$ \\
    \hline
  \end{tabular}
\end{table}

Molecular dynamics simulations were performed using an integration time step of 10 ps. The simulated depositions were allowed to progress for 1 ms ($10^8$ steps) and configurations were stored every $10^4$ steps. The computational resources required for these simulations is a strong function of $\lambda_D$. Using Lawrence Livermore National Laboratory's Quartz supercomputer, with 2 18-core Intel Xeon E5-2695 processors per node, $\lambda_D$=0.5 nm simulations were completed in $\sim$17 hours of wall time on 8 nodes. Simulations  with the longest Debye length, $\lambda_D$=50.0 nm, required $\sim$14 days of wall time on 25 nodes.  

\section{Results and discussion}

Here we present an overview of the four model scenarios described in Table~\ref{tbl:fixes}. Each scenario was run with three Debye lengths and under three electric fields for a total of 36 EPD simulations. This simulation set provides insights into various non-ideal deposition behaviour that can arise in real EPD systems, but are not accounted for in the standard DLVO particle-based model, termed here as \emph{standard/mobile}. We present a quantitative overview of the four cases, discussing pertinent empirical systems that may exhibit such deposition phenomena. We then walk through our  analysis of density profiles and first-layer particle ordering and present quantitative comparisons between the EPD model types across our simulation set. 

\begin{figure}[h!]
\includegraphics[width=3.0in]{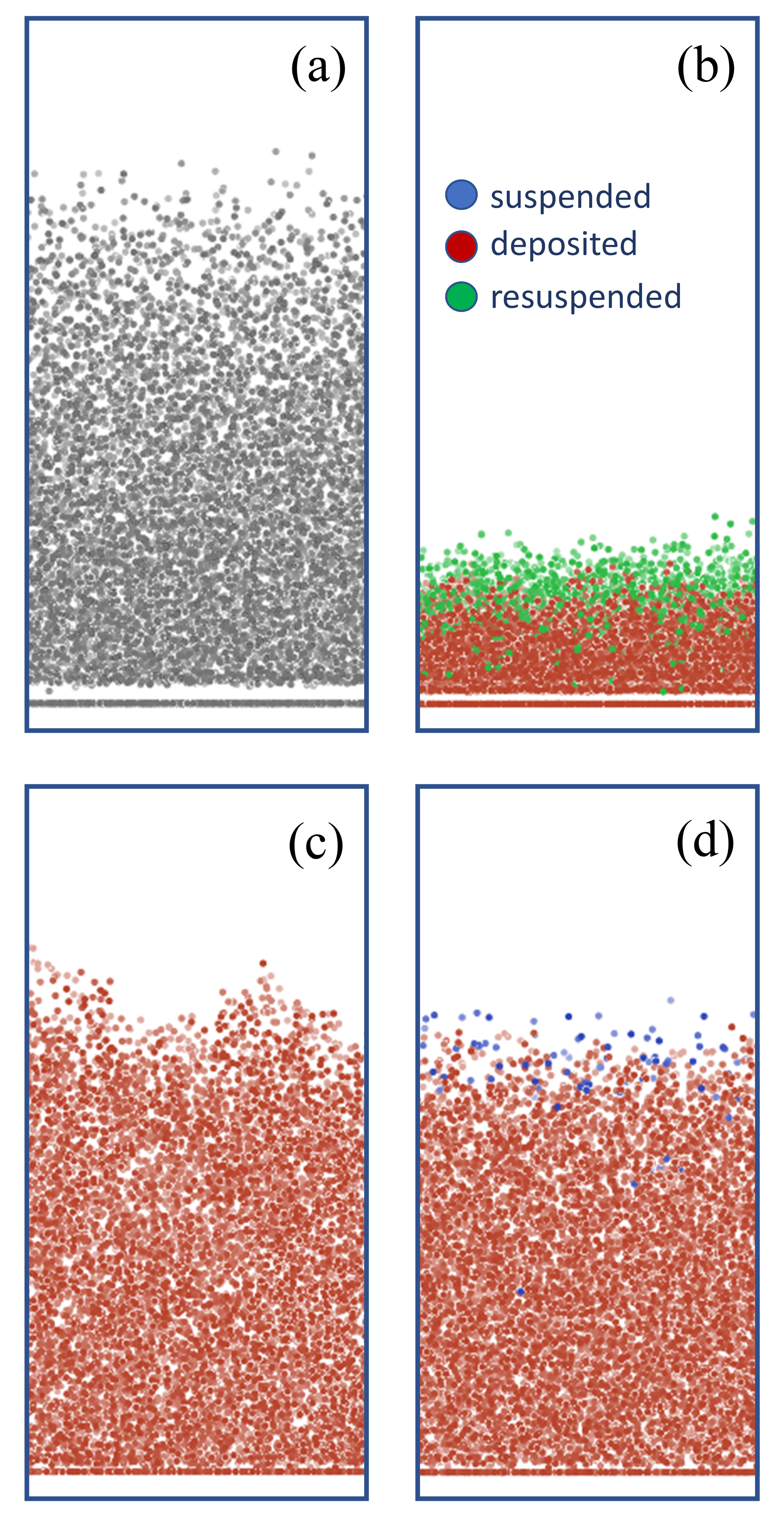}
\caption{\label{fig:fixes} Final frames of the four model types: (a) \textit{standard/mobile} model in which all particles interact solely through the DLVO pairwise potential. (b) \textit{decharge/mobile} in which particles lose their charge when depositing onto the electrode or other deposited particles but stay mobile. Resuspended (green) and deposited (red) particles interact with the uncharged DLVO potential. (c) \textit{decharge/immobile}, a discharging model in which deposited particles are immobilized. (d) \textit{standard/immobile} model in which deposited particles are immobilized but retain their charge.}
\end{figure}

Figure \ref{fig:fixes} shows snapshots of simulated EPD after 1~ms of simulation time for each of these extremes of possible physical EPD phenomena. These representative simulations are all performed with  $\lambda_D$ = 5.0 nm and a large, though experimentally accessible, electric field of $E_z$ = 250 kV/m. Panel (a) shows the \textit{standard}/\textit{mobile} approach, which simulates deposition using consistent application of DLVO interaction potentials and physics model detailed in above and in Reference \citenum{gieraMesoscaleParticleBasedModel2017}. We represent the colloid particles in Figure \ref{fig:fixes}a with gray color to emphasize that none of the new deposition modes introduced in this work are implemented in the \textit{standard}/\textit{mobile} simulations: The physics model for bulk, deposited, and re-suspended colloid particles is constant. 

In Figure \ref{fig:fixes}a the first deposition layer is isolated from the rest of the deposit, separation due to the close-packing at the planar electrode and a combination of electrostatic repulsion and excluded volume due to the colloids' size. Subsequent layers of the deposited colloid quickly become less ordered and the density of the deposit  decreases toward the bulk. For DLVO colloidal suspensions with large Debye lengths, thick and/or dense deposits are unlikely to form where mobile particles must overcome a repulsive energy barrier from already-deposited particles. Indeed, deposits under these conditions are known to resuspend in the absence of the electric field~\cite{PhysRevMaterials.4.075802,Giera_2018}. These empirical observations and complementary simulation results fit within the standard DLVO paradigm. However, in some sense, they are counter-intuitive considering how researchers routinely create dense, lasting deposits under a wide variety of conditions. The implementation of DLVO theory as a conventional pairwise additive molecular dynamics force field is too limited a framework to capture many EPD approaches. We present three additional EPD simulation scenarios with physics models that adapt to particles' local environments on-the-fly in an effort to explore new physical effects beyond conventional particle-based simulation methods.

Figure \ref{fig:fixes}(b) shows the \textit{decharge}/\textit{mobile} approach. Several features of this deposit differ significantly from (a). The excluded region between the first deposition layer and bulk deposit is smaller due to the removal of particle-particle electrostatic repulsion in the deposited material. The lack of this repulsion also results in a much denser deposit, $\sim25\%$ as thick as the \textit{standard}/\textit{mobile} deposit in (a). In (b) we color particles according to whether they are deposited (touching the electrode or within $r_{\rm infect}$ of deposited particles) or resuspended (once-deposited particles that migrated beyond $r_{\rm cure}$ of the deposit). Experimental systems that exhibit this behavior include examples in which functionalized or ligand-coated particle and electrode surfaces can exchange charge via acid-base interactions~\cite{chen_modulation_2021}.

Snapshots of the \textit{immobile} simulations, shown in Figure \ref{fig:fixes}(c) and (d), appear similar. In both snapshots, the separation of the first deposited layer of particles is less apparent than in either of the \textit{mobile} simulations and the thickness of the two deposits is similar. The most obvious difference between the two snapshots is the persistence of some suspended particles in (d), the \textit{standard}/\textit{immobile} case, since the repulsive electrostatic forces are still present after particles are tagged as deposited and made immobile. The \textit{decharge}/\textit{immobile} case in Figure~\ref{fig:fixes}(c) is representative of ligand-free metallic particle suspensions~\cite{doi:10.1021/acs.chemrev.6b00468}. Another example of the \textit{decharge}/\textit{immobile} case is the absorption of Pt nanoparticles on oxidic supports, where \citeauthor{marzun_adsorption_2014} demonstrated that charge transfer could be observed by monitoring surface charge density and zeta potential $\zeta$.~\cite{marzun_adsorption_2014} In this example the \textit{decharge} scenario would be applicable even though there was no externally applied electric field. Figure~\ref{fig:fixes}(d), characterized by the \textit{standard}/\textit{immobile} scenario, can occur when a gelation agent immobilizes particles at the electrode. Application drivers for this design include sensing, energy conversion, and catalysis.~\cite{geng_electrochemical_2023,dickerson_electrophoretic_2012} Additionally, studying the differences between \textit{decharge}/\textit{immobile} and \textit{standard}/\textit{immobile} scenarios may provide insights into how the nature of particle-electrode bonding affects the emergent deposit.

\subsection{Quantifying simulated EPD}

\begin{figure}[h]
\includegraphics[width=6in]{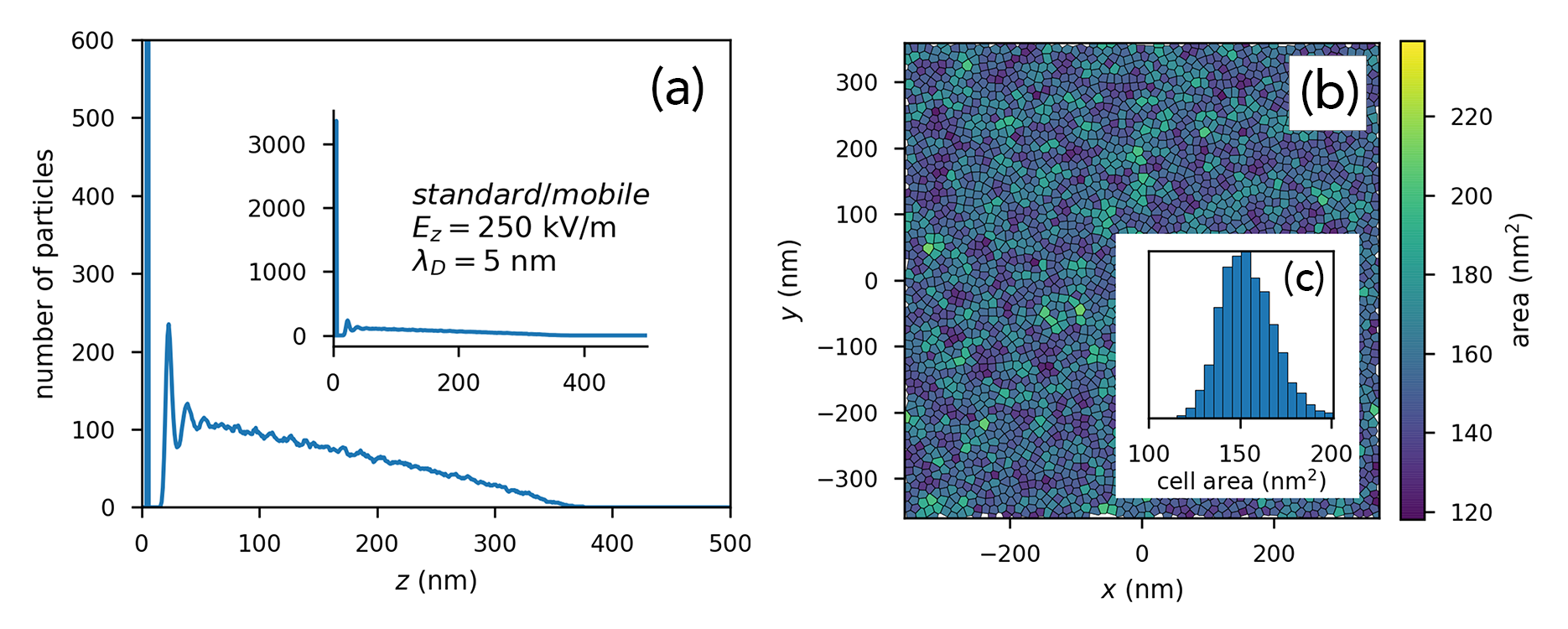}
\caption{\label{fig:overview} (a) Density profile and (b) Voronoi diagram describing the deposit resulting from simulated EPD deposition in the \textit{standard}/\textit{mobile} case where $E_z=250$ kV/m and $\lambda_D$ = 5 nm. The inset (c) is a probability histogram of Voronoi cell areas.}
\end{figure}

Our quantitative analysis of the 36 EPD simulation set focuses on two main features: the density profile and the spatial arrangement of the first layer of colloid particles at the planar electrode. Figure \ref{fig:overview} shows representative data for the case of \textit{standard}/\textit{mobile}, $\lambda_D$=\SI{5}{nm}, and $E_z$=250 kV/m. 

The density profiles in panel (a) are the average of the final thousand stored configurations of the 1 ms MD trajectory. We note that additional sampling beyond the final frame of \textit{immobile} depositions does not reduce statistical noise in the deposit. For all density profiles the vertical axis is shortened, truncating the peak that corresponds to the first layer of deposited particles to enable better inspection of the bulk deposit. The inset of Figure \ref{fig:overview}a shows same density profile with full range of the vertical axis. 

Figure \ref{fig:overview}(b) shows the first layer of deposited particles as a Voronoi diagram. Construction of Voronoi diagrams in this work is as follows: We isolate the first deposited layer by extracting all particles $z_i<2.2r$ from the last stored configuration of a simulation trajectory and projecting them onto a plane by removing the $z$ Cartesian coordinate. For the given set of particles at the electrode $N_1,...,N_p$, each particle $N_i$ has a corresponding Voronoi cell, $V_i$, which contains every point in the plane whose euclidean distance to $N_i$ is less than the distance to any other particle at the electrode. Each Voronoi cell represents one particle. In this work, Voronoi diagrams representing the first deposition layer are calculated and visualized using an in-house customization of the freud Python library, where the color of each Voronoi cell corresponds to its area.~\cite{freud2019,freud2020} The inset Figure \ref{fig:overview}(c) is a histogram of the Voronoi cell sizes, which represents the electrode surface area occupied by the deposited colloid particles. Voronoi analysis has previously been applied to analyze scanning electron microscopy images of EPD monolayer films where the particles were characterized by the number of sides on its Voronoi cell, corresponding to the number of nearest neighbor particles.~\cite{krejci_using_2013} Voronoi analysis presents a straightforward way survey surface coverage by EPD and to directly and quantitatively compare particle-based simulations and experiment.

\subsection{Behavior of deposition models}

\begin{figure}[h]
\includegraphics[width=5in]{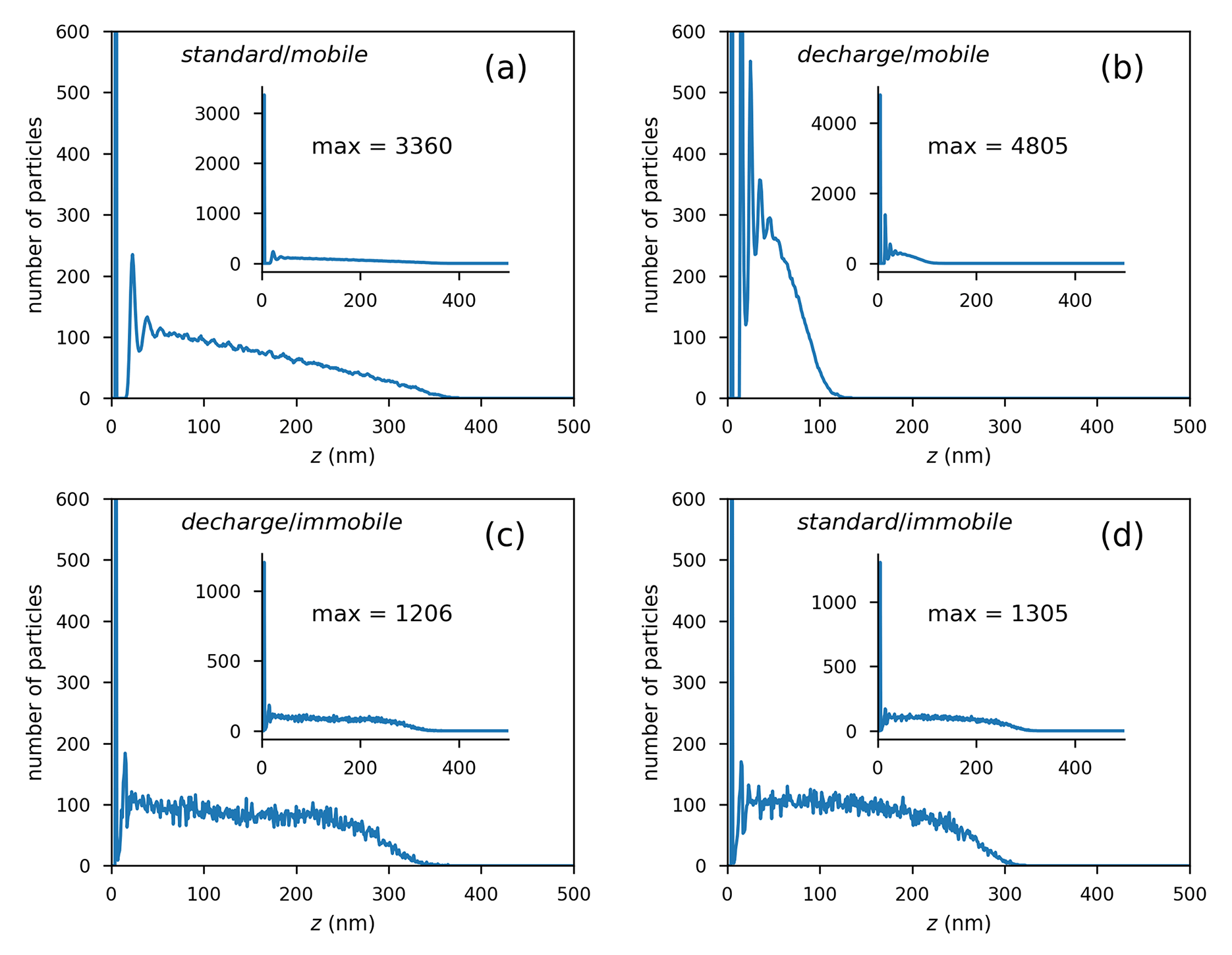}
\caption{\label{fig:densProf4} Deposit density profiles for the four simulation cases: (a) \textit{standard}/\textit{mobile} (b) \textit{decharge}/\textit{mobile} (c) \textit{decharge}/\textit{immobile}, and (d) \textit{standard}/\textit{immobile}.  For all cases $\lambda_D$ = 5 nm and $E_z$ = 250 kV/m. Inset plots show the full $y$-axis and the annotation (``max = $N$'') is the height of the first peak, which roughly corresponds to the number of particles at the electrode.} 
\end{figure}

The density profiles in Figure \ref{fig:densProf4} provide direct, quantitative insight into the simulated deposits.  In (a), the \textit{standard}/\textit{mobile} simulation, three layers of decreasing ordering are visible before the particle density decays to a smooth, monotonically decreasing profile, ending at $\sim$ \SI{300}{nm} from the electrode. The density profile (b), corresponding to the \textit{decharge}/\textit{mobile} reflects the changes to the deposited particles' physics model: deactivation of the electrostatic repulsion between the particles. Removing this repulsive component of the DLVO model decouples the ordering of the deposit from $\lambda_D$, enabling the deposited particles to pack more closely together. This results in much more ordering, similar to hard sphere packing, and 5 well-defined peaks that correspond to the first 5 layers of the deposit.

The \textit{immobile} approaches are shown in Figure \ref{fig:densProf4}(c) and (d). We again note that, since deposited particles are made completely immobile, statistical noise may not be reduced by time-averaging snapshots. This results in significantly noisier density profiles for the \textit{immobile} cases. To a first approximation, (c) and (d) look similar, with high density layers at the electrode followed by significantly less ordered second layers. The bulk of each deposit has a flat profile relative to the \textit{mobile} cases.

\clearpage

\begin{figure}[h]
\includegraphics[width=6.5in]{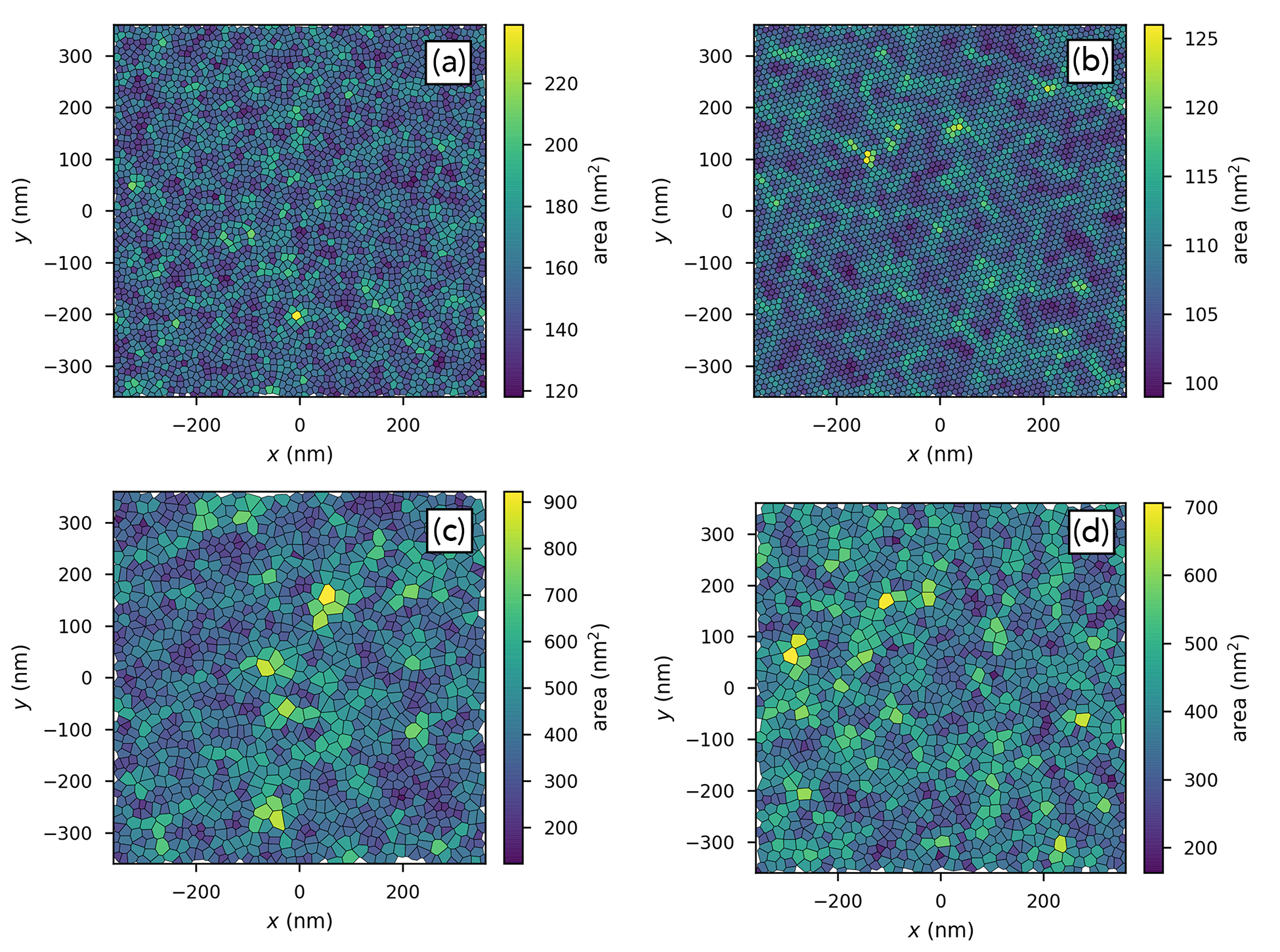}
\caption{\label{fig:voro4} Voronoi diagrams for each deposition model: (a) \textit{standard}/\textit{mobile} (b) \textit{decharge}/\textit{mobile} (c) \textit{decharge}/\textit{immobile}, and (d) \textit{standard}/\textit{immobile}. For all cases $\lambda_D$ = 5 nm and $E_z$ = 250 kV/m.}
\end{figure}

\begin{figure}[h]
\includegraphics[width=3.in]{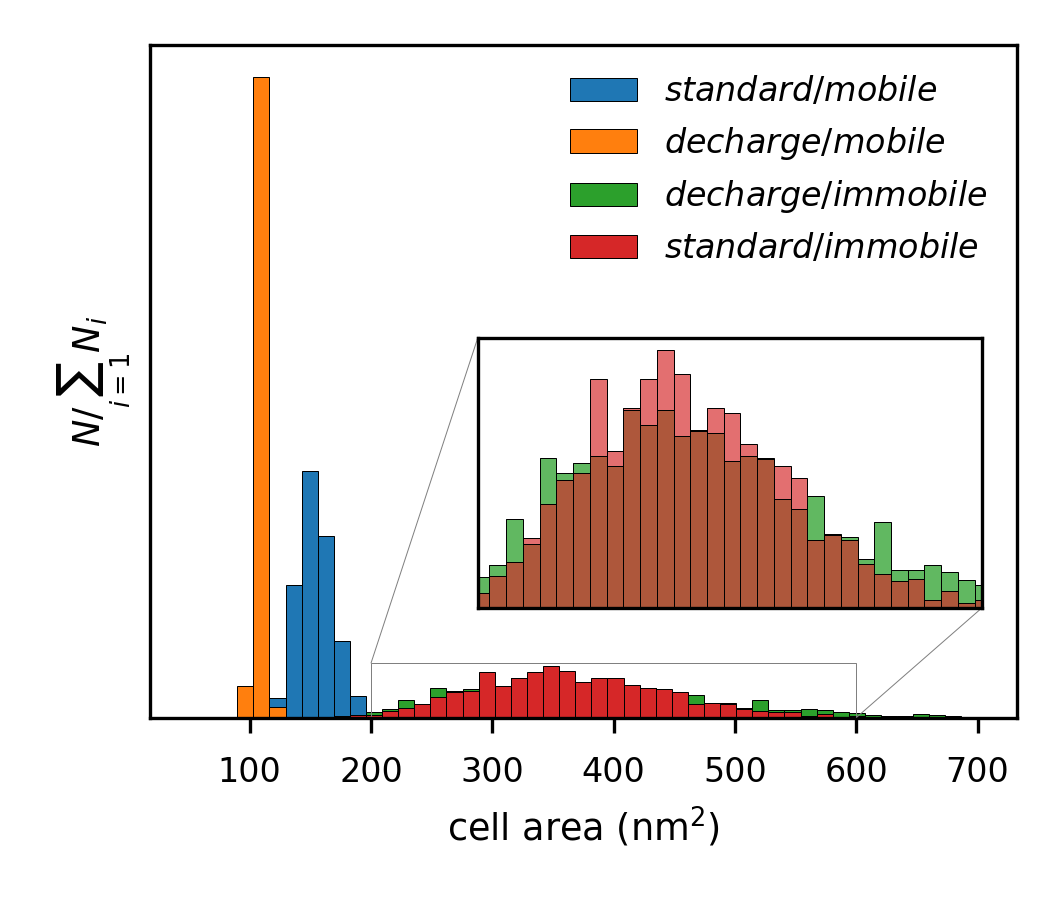}
\caption{\label{fig:voroh4} Voronoi cell area histograms corresponding to the Voronoi diagrams of the four deposition models in in Figure \ref{fig:voro4}. Histograms are normalized to unit total area. The inset image magnifies the \textit{standard/immobile} and \textit{decharge/immobile} histograms and shows the bars as semitransparent to reveal differences between the two results.}
\end{figure}

Figure \ref{fig:voro4} shows representative Voronoi diagrams of the first layer of deposited particles at the electrode for each of the four deposition models, with $\lambda_D$ = 5.0 nm and $E_z$ = 250 kV/m for all cases. The area of each Voronoi cell corresponds to the area of the electrode occupied by a given particle; the number of cells corresponding to the number of particles at the electrode. Panel (a) is the \textit{standard}/\textit{mobile} case, which corresponds to simulations presented in earlier work.~\cite{gieraMesoscaleParticleBasedModel2017} Implementation of the \textit{decharge} model, Figure \ref{fig:voro4}(b), results in a more densely packed layer of particles at the electrode, agreeing with data from the density profiles in Figure \ref{fig:densProf4}. Histograms of these Voronoi cell areas are overlaid and shown in Figure \ref{fig:voroh4}, where the \textit{standard}/\textit{mobile} distribution is centered at \SI{150}{nm^2}. The \textit{decharge}/\textit{mobile} particles are more closely packed, with a considerably smaller variance and an average area of \SI{108}{nm^2}. When projecting the first deposition layer onto a plane, this corresponds to 73\% of the particles' area covering the electrode surface versus 51\% in the \textit{standard}/\textit{mobile} case and 90.7\% for the case of a hexagonal lattice: the ideal, highest-density configuration for packing circles in a plane.~\cite{fejesUeberDichtesteKugellagerung1942}   

The \textit{immobile} cases, shown in Figure \ref{fig:voroh4}(c) and (d), show significantly less dense packing at the electrode. Although the electrostatics of the deposits formed in the \textit{decharge}/\textit{immobile} (c) and \textit{standard}/\textit{immobile} (d) cases differ substantially, the average Voronoi cell areas are within $\sim$4\%. This indicates that the separation and ordering of the particles at the electrode in these simulations is a strong function of particle deposition parameters, $r_\text{sticky}$ and $r_\text{infect}$, and not dominated by electrostatic repulsion as in the \textit{mobile} cases. This similar behavior at the electrode is due to the suspended particles descending with identical physics until they interact with the electrode, then becoming `deposited' and made immobile. In the first layer changes in the systems' electrostatics (i.e. $U_\text{Yuk}$) are relatively small. As the deposited layers grow, differences in the deposits' electrostatics increase which results in the differences observed in the \textit{immobile} density profiles in Figure \ref{fig:densProf4}(c) and (d). The \textit{standard} simulation, which retains DLVO behavior throughout the deposit, results in additional suspension-deposit particle-particle repulsion and a slightly more well-ordered and more dense deposition. This repulsion is seen qualitatively in Figure \ref{fig:fixes}(d) where a few particles remain suspended (blue) above the deposit after the simulation has concluded. Other instances of this behavior are observed within the selected parameter sweep data presented in the following sections.

\clearpage
\subsection{Varying electric field for fixed Debye length of 5 nm}

\begin{figure}[h]
\includegraphics[width=6.5in]{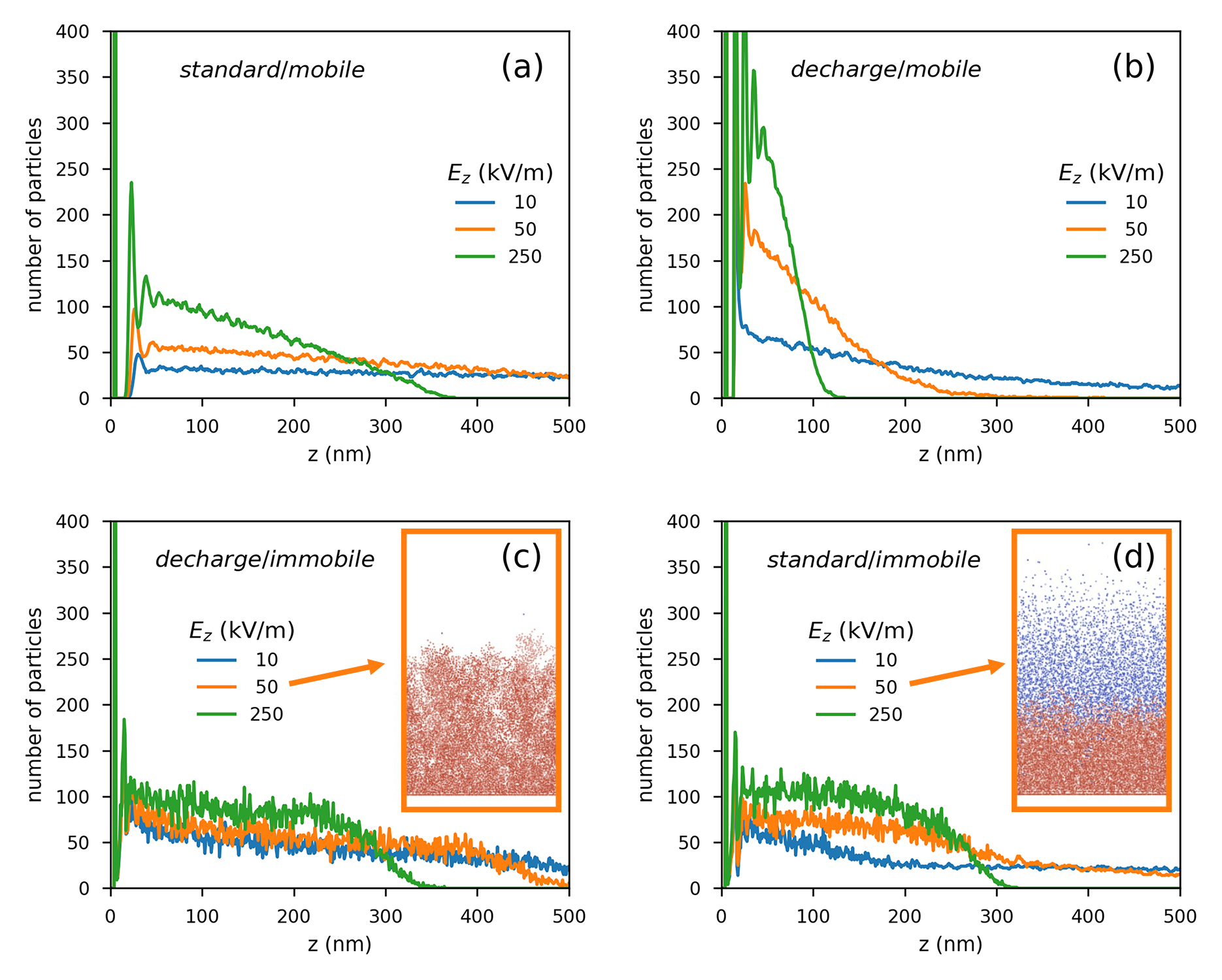}
\caption{\label{fig:efield} Deposit density profiles for the four models. Each panel shows the resulting deposit under applied fields of 10 (blue), 50 (orange), and 250 (green) kV/m. The insets in (c) and (d) are $E_z=50$ kV/m simulations snapshots obtained at $t=1$ ms that reveal deposited (red)/suspended (blue) population differences.}
\end{figure}

Figure \ref{fig:efield} shows density profiles, where each panel contains data for a single deposition model. The three curves in each panel correspond to different field strengths $E_z$: 10 (blue), 50 (orange), and 250 (green) kV/m. The Debye length, $\lambda_D$, is 5 nm for all 12 simulations shown in Figure \ref{fig:efield}. For all deposition models high field strength results in a more dense deposit, although the deposits vary greatly between deposition models.  The \textit{mobile} deposition models, panels (a) and (b), show similar trends, with higher packing within the \textit{decharge} case due to the loss of interparticle electrostatic repulsion upon deposition. Although not readily apparent in the density profiles, all \textit{decharge}/\textit{mobile} cases have a significant layer of `resuspended' particles adjacent to the particles tagged as `deposited,' as seen in Figure \ref{fig:fixes}(b). This layer of particles, electrostatically perturbed by contact with the electrode but suspended in solution, appears to act as a passivation layer, particularly at lower fields, preventing suspended particles in the bulk from reaching the electrode (or deposited particles in intimate contact with the electrode) and losing their electrostatic character. This effect may be made more evident by removing the `resuspension' scheme (see Figure \ref{fig:fix-toons}(c)). With \texttt{fix cure} disabled, deposited particles that diffuse away from the electrode move through the bulk suspended colloid, removing the electrostatic repulsion from all suspended particles they encounter and all \textit{decharge}/\textit{mobile} simulations result in a dense-packed deposit. Using the terminology of our LAMMPS implementation, \texttt{fix cure} is a first-order approximation to prevent this nonphysical runaway electrostatic deactivation by \texttt{fix infect}.

The \textit{immobile} cases in panels (c) and (d) appear similar to each other when compared to the \textit{mobile} cases. Upon close inspection \textit{decharge}/\textit{immobile} and \textit{standard}/\textit{immobile} density profiles show differences that result from the deposits' dissimilar electrostatic behavior. Since particles are made immobile upon deposition these cases isolate the effect of deposit-suspension repulsion. This repulsion affects the ordering of the deposits in the \textit{standard} case, resulting in slightly higher density packing near the electrode. In the \textit{decharge} case the lack of deposit-suspension interaction permits a stochastic deposition; this lack of ordering results in the observed lower densities. During \textit{immobile} in silico deposition, the fraction of particles tagged as `deposited,' varies more dramatically in simulations where  deposit-suspension electrostatic repulsion becomes stronger than the applied field, $E_z<250$k V/m. The density profiles in (c) and (d) corresponding to $E_z=50$k V/m (orange curves) show more interesting differences. In (c), the \textit{decharge} case, the rapid density decrease between 400 and 500 nm indicates the upper boundary of the deposit. A similar decrease is not observed in (d), although a slight decrease at around 250-300 nm occurs before the density plateaus. The inset images in panels (c) and (d), simulation snapshots corresponding to $E_z=50$k V/m, reveal the origin of these density features: deposit-suspension repulsion in (d) limits the extent of deposition, even in our \textit{immobilize} protocol, which implements the most liberal case of immobile particle deposition. 


\clearpage

\begin{figure}[h]
\includegraphics[width=6.5in]{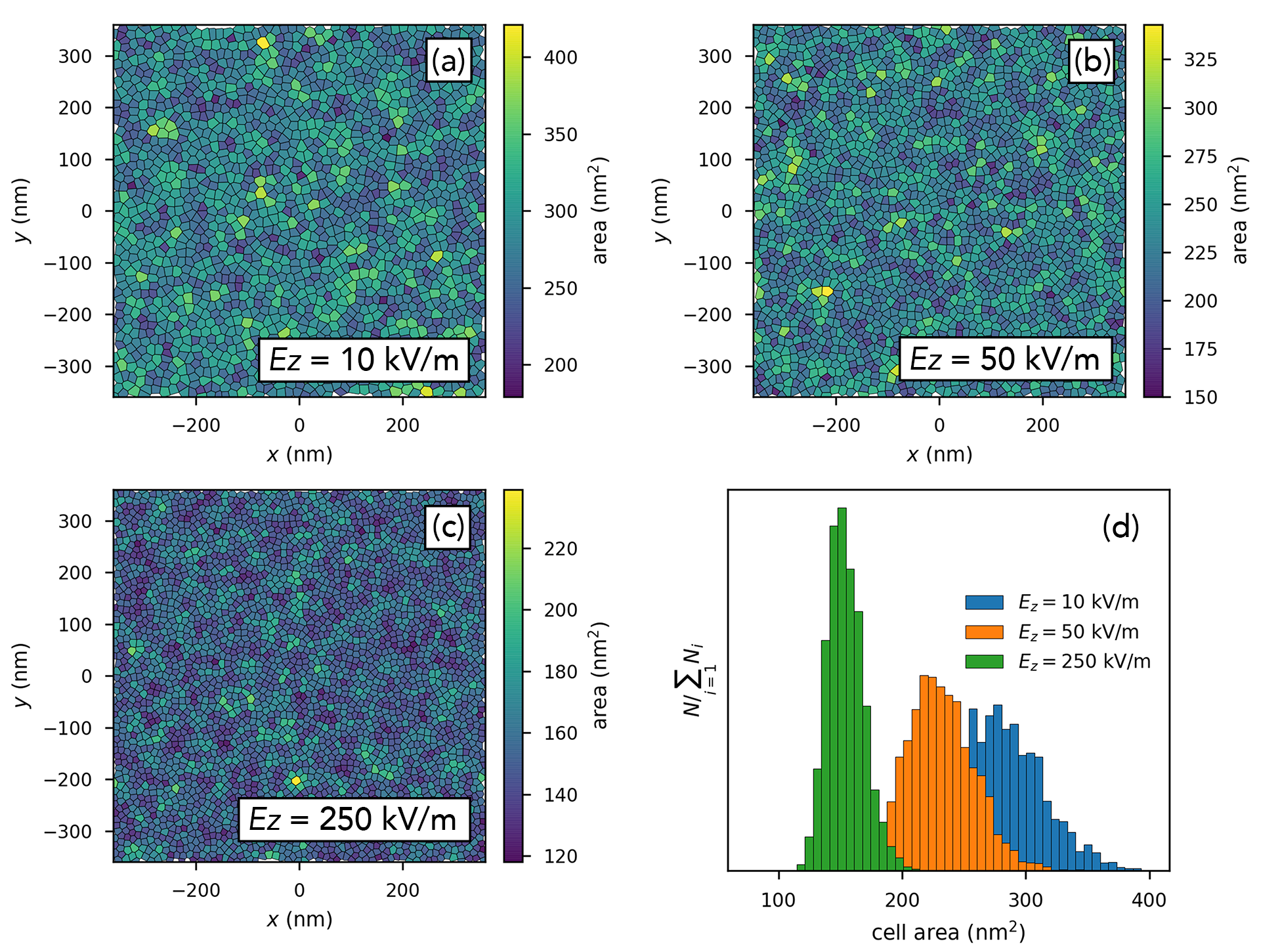}
\caption{\label{fig:voro} Effect of varying $E_z$ on first deposition layer: (a) $E_z$ = 10 (b) $E_z$ = 50 (c) $E_z$ = 250 kV/m. The deposition model is \textit{standard}/\textit{mobile} and $\lambda_D$ = 5 nm for all panels.} Panel (d) shows distributions of the Voronoi areas for panels (a) - (c). 
\end{figure}

Figure \ref{fig:voro} shows Voronoi diagrams for the \textit{standard}/\textit{mobile} simulations for selected $E_z$. For this deposition model colloid spacing at the electrode decreases with increasing field, showcasing the interplay between interparticle repulsion and $E_z$, the primary driving force of particle deposition at the electrode. Histograms of the Voronoi cell areas in panel (d) are an alternate presentation of these areas; more succinct but sacrificing the overview of spatial homogeneity of the particles' effective areas. Approximating each cell as a circle estimates the effective radius of particles at the surface, 9.4, 8.6, and 7.0 nm for the 10, 50, and 250k V/m fields, respectively, as compared 5.25 nm for the case of ideal packing of circles in a plane. This phenomenon of more dense and clustered deposits at higher field strength has been experimentally observed for Pt nanoparticle deposition.~\cite{cernohorsky_insight_2016}

\clearpage

\subsection{Varying Debye length for fixed electric field of 250 kV/m}

\begin{figure}[h]
\includegraphics[width=6.5in]{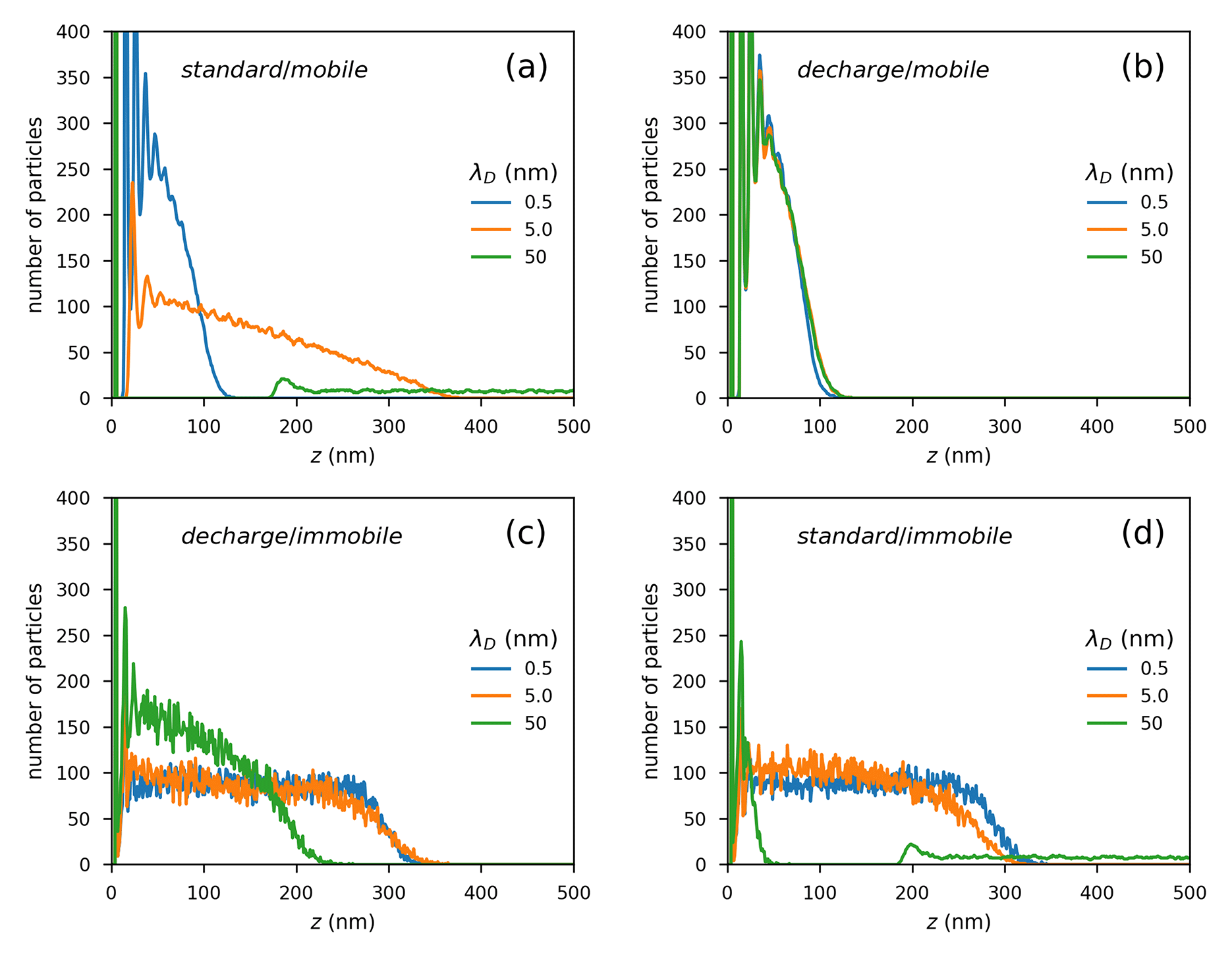}
\caption{\label{fig:lambda} Deposit density profiles for the four simulation scenarios. Each panel shows the resulting deposits for the given scenario and each colored curve corresponds to a different value of $\lambda_D$: 0.5 (blue), 5.0 (orange), and 50 (green) nm. $E_z=250$ kV/m for all simulations.}
\end{figure}

Figure \ref{fig:lambda} shows density profiles that result from varying the Debye length ($\lambda_D$) of the colloid particles. Similar to Figure \ref{fig:efield}, each panel represents one of the four deposition models. The three curves in each panel correspond to different $\lambda_D$: 0.5 nm (blue), 5 nm (orange), and 50 nm (green). In panel (a), the \textit{standard}/{mobile} case, the shortest $\lambda_D$ results in the most densely packed deposit, only about 100 nm thick. The $\lambda_D=5$ nm simulation shows the characteristic close-packed and well-ordered first deposition layers followed by linear decrease in deposit density. The large 50 nm $\lambda_D$ simulation shows the most interesting result, where a high-density first layer is present, followed by region of near-zero density that extends to $\sim180$ nm from the electrode surface. A small density peak is present after this excluded region, followed by a low-density colloid suspension. 

All three density profiles in panel (b) look identical because the \textit{decharge}/\textit{mobile} case removes interparticle electrostatics, thus removing the $\lambda_D$ differences introduced in this parameter study. Deposited particles in this case effectively behave as $\lambda_D=0$. This explains the similarity between the $\lambda_D=0.5$ nm curve in panel (a) and the curves in (b) since the range of electrostatic repulsion in the $\lambda_D$ simulations is significantly smaller than the particles' radii. 

The shorter $\lambda_D$ \textit{immobile} cases in (c) and (d) look similar aside from the slightly more dense deposit in the \textit{standard}/\textit{immobile} case due to additional ordering of the descending particles by particle-particle repulsion. Conversely, the deposits resulting from the $\lambda_D=50$ nm cases look very different. In the \textit{decharge} case, the extended electrostatics in the suspension results in more ordering during the deposition and a more dense deposit than the $\lambda_D=0.5$ and 5 nm cases, a trend that suggests a design strategy for higher-density EPD deposits within this physical regime. In panel (d), the \textit{standard}/\textit{immobile} case, the $\lambda_D=50$ nm deposit resembles the $\lambda_D=50$ nm density profile in (a), where the first layer of deposit repels the suspended bulk colloid. In the \textit{immobile} case, this results in a suspended colloid regime separated from the electrode at that same distance as in the analogous \textit{mobile} simulation.

\section{Conclusions}

The particle-based modeling approaches presented here represent extensions beyond standard DLVO-based implementations for EPD simulation. As such, they pave the way for exploring the interplay between electrode and colloidal suspension properties and EPD operating conditions that may result in non-DLVO phenomena. Specifically, we model four types of deposition in which particles (1) keep or (2) lose their charge upon deposition and/or (3) remain mobile or (4) lose mobility upon deposition. We simulate parameter sweeps of these scenarios and compare resulting particle density profiles throughout the deposit and ordering at the electrode. Compared to the \emph{standard/mobile} case, \emph{decharge/mobile} deposits pack more densely in the absence of electrostatic repulsion between deposited and (re)suspended particles. This reveals the impact of Debye length on the structure and density of EPD deposits: If $\lambda_{D}$ is too large, multilayer deposits may not form due to repulsion from deposited particles. Deposit density increases with electric field but is, in all cases, less dense than deposits formed in `mobile' particle scenarios.

This extensible and versatile EPD modeling framework is an initial step toward expanding particle-based EPD simulation to capture many more physical systems. Modifications to any of the four EPD model types shown here are straightforward to implement. For instance, as with previous particle-based models, the functional form of the interaction potentials among particles and electric field are tunable and, with this work, dynamic. Furthermore, the cutoff distances used to specify whether particles are deposited or suspended can be modified accordingly to match or approximate the behaviour of any physical system of interest. Simulations can be configured to account for multiple deposition scenarios, e.g. DLVO particles in which a fraction irreversibly adsorb to the surface. Since this framework allows for conditional changes to any property, it is possible to simulate colloids that partially decharge or exhibit reduced mobility near the electrode or deposit.  

Though our proof-of-concept simulation set focuses on the extremes of EPD physical phenomena, it is possible to interpolate insights pertaining to more complicated, non-ideal systems commonly encountered in the EPD research community. In cases where multiple deposition effects may be at play and/or for EPD conditions not represented here, the model readily can be tailored to simulate a wide set of experimental systems. Thus, it offers new ways of guiding the experimental design of particle deposition beyond what was possible with previous modeling capabilities.  Overall, the model provides an unprecedented quantitative glimpse into the EPD process for \emph{both} measurable and (presently) empirically inaccessible EPD phenomena, e.g. particle migration within deposits. Considering the various falsifiable aspects of the model, e.g. bulk particle trajectories, surface characterization, etc., model validation and complementary model improvement strategies can be achieved through a tight coupling between experiment and theory-based research.


\begin{acknowledgement}
This work was performed in part under the auspices of the U.S. Department of Energy by Lawrence Livermore National Laboratory under contract DE-AC52-07-NA27344, release number LLNL-JRNL-849162.
We gratefully acknowledge the German Research Foundation (DFG) for their financial support under the project number BA 3580/24-1 and B.G. thanks DFG for the Mercator Fellowship.

\end{acknowledgement}

\begin{suppinfo}

Additional details regarding Yukawa potential cutoff distances, source code for the adaptive physics models, and a representative animation showing the four EPD simulation scenarios studied are included in addition to the main text.

\end{suppinfo}

\bibliography{achemso-demo}

\end{document}




\clearpage

\section{Yukawa potential cutoffs}

\begin{figure}[h]
\includegraphics[width=5in]{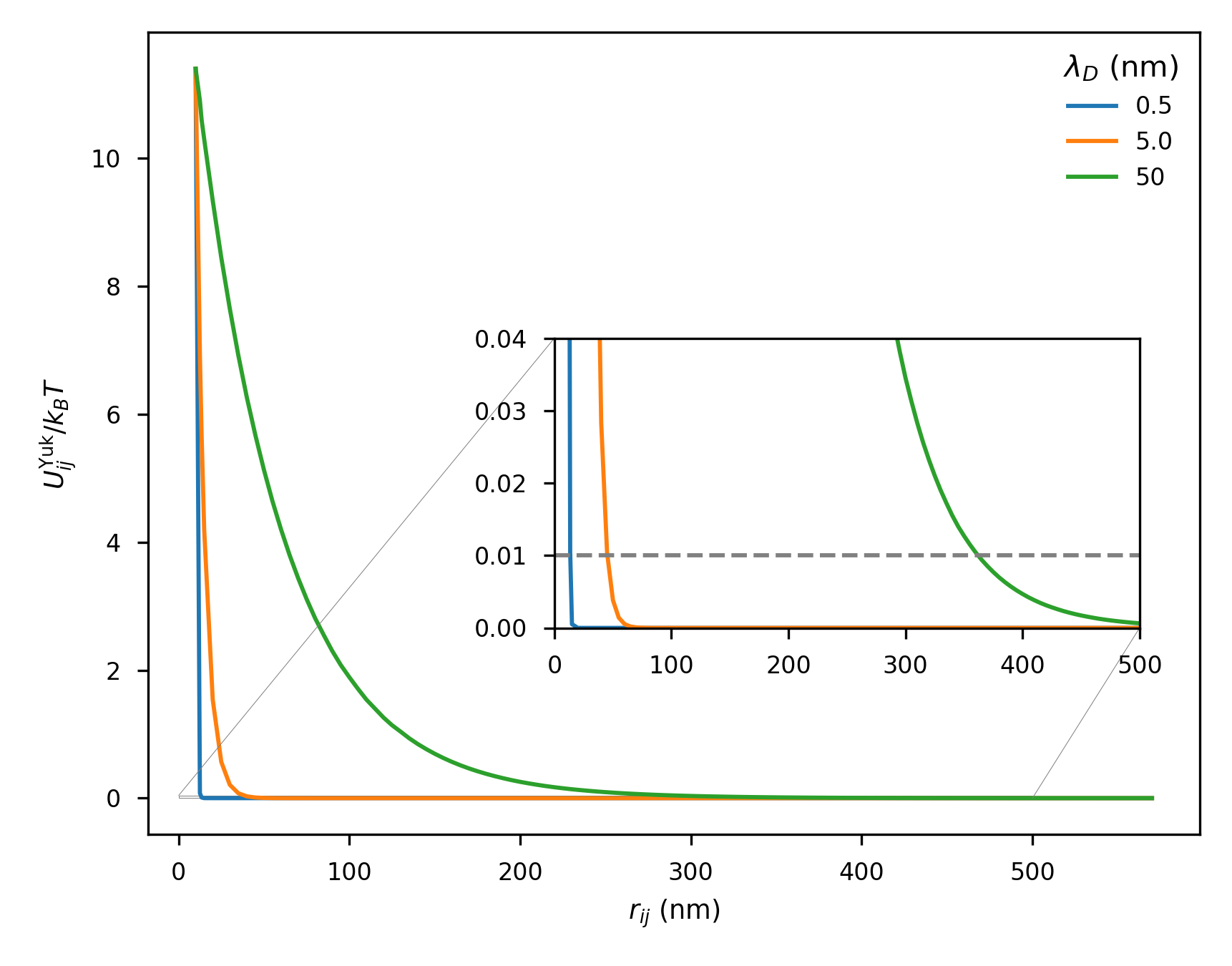}
\caption{\label{fig:yukCut} The Yukawa potential's contribution to pairwise interaction energy for simulated colloids. $U^{\text{Yuk}}_{ij}$ is shown in units of $k_BT$ versus the pairwise distance, $r_{ij}$. The inset shows the region of $0-4\%$ $k_BT$ and the dashed line is included to guide the eye to 1\% $k_BT$, which defines the $U^\text{Yuk}$ cutoffs used in this work.} 
\end{figure}

\clearpage

\begin{acknowledgement}
This work was performed in part under the auspices of the U.S. Department of Energy by Lawrence Livermore National Laboratory under contract DE-AC52-07-NA27344, release number LLNL-JRNL-849162. We gratefully acknowledge the German Research Foundation (DFG) for their financial support under the project number BA 3580/24-1 and B.G. thanks DFG for the Mercator Fellowship.

\end{acknowledgement}

